# Line Outage Detection and Localization via Synchrophasor Measurement

Xianda Deng[2], Desong Bian[1], Di Shi[1], Wenxuan Yao[2,3], Zhihao Jiang[2], Yilu Liu[2,3]
[1] Global Energy Interconnection Research Institute North America
[2] University of Tennessee Knoxville, TN, [3] Oak Ridge National Lab

xdeng6@vols.utk.edu, desong.bian@geirina.net, di.shi@geirina.net,
wyao3@ vols.utk.edu, zjiang13@vols.utk.edu, liu@utk.edu

*Abstract*—Since transmission lines are crucial links in the power system, one line outage event may bring about interruption or even cascading failure of the power system. If a quick and accurate line outage detection and localization can be achieved, the system operator can take necessary actions in time to mitigate the negative impact. Therefore, the objective of this paper is to study a method for line outage detection and localization via synchrophasor measurements. The density of deployed Phasor Measurement Units (PMUs) is increasing recently, which greatly improves the visibility of the power grid. Taking advantage of the high-resolution synchrophasor data, the proposed method utilizes frequency measurement for line outage detection and power change for localization. The procedure of the proposed method is given. Compared with conventional methods, it does not require the pre-knowledge on the system. Simulation study validates the effectiveness of the proposed method.

## I. INTRODUCTION

The detection and localization of transmission line outage in power system is of great significance for the system operators to take prompt action to avoid the widespread damage and maintain the reliability of power supply[1]-[5]. Most of current methods rely on angle data along with network susceptance matrix to calculate power injection change, which is a high computation burden and requires the information of system parameter [2], [6]-[11]. The method using PMU angle data and network susceptance was originally proposed in [2] and pre- and post-outage power flow were calculated to match the measured event. Later, compressive sensing and global optimization techniques are proposed to improve the method in [7] and [8]. A general Bayesian criterion was employed to handle the uncertainty issue of PMU data in [9]. Different new schema and frame are developed to deal with bad PMU measurements in [10] and [11]. However, the system parameter may not be available all the time due to strict security concerns.

With the rapid transformation from traditional power system into the smart grid, there are various types of novel applications involved into the system [12]-[15]. These smart grid applications are relying on the high quality synchronized data and advanced communication network infrastructure [16]-[24], such as synchrophasor measurements, advanced metering infrastructure, and home energy management system. Nowadays, the density of synchrophasor is increasing dramatically to observe the dynamic behavior of the system following a contingency, which gives the unpreceded insights to the system [23]-[28]. For example, there are 114 and 238 PMUs installed at Jiangsu power grid, respectively. As shown in Fig. 1, the density of distributed PMU is significant high, which covers all 500kV transmission lines and parts of 220 kV transmission lines with the reporting rate 25 Hz. To utilize the PMU data and achieve wide area monitoring purpose, a PMU based situational awareness data analytics platform has been developed by Global Energy Interconnection Research Institute North America (GEIRINA) [29]-[33]. The PMU based platform collects synchrophasor measurements with massive channels in real time from Jiangsu power grids and processes large amount of data, which can be affected by latency from PMU device or communication network. The platform not only incorporates event detection application developed by GEIRINA, but also provide interface for event detection applications from third party. The density synchrophasor measurement brings the opportunity to detect the line outage location and locate the fault location without knowing the system parameters. The line outage detection approach introduced by [34] was employed in the PMU based platform and the reported line outage locations have significant deviation from actual line outage location. Similar phenomenon was also founded in simulation cases in New England ISO (ISO-NE) and Tennessee Valley Authority (TVA) systems.

To address the issue mentioned above, this paper focuses on the method for line outage detection and localization via synchrophasor measurement. First, the line outage is detected via employing low pass filter and peak detector on synchronized frequency measurements. Once a line outage event is triggered, the location of the fault line will be pinpointed using power flow change. The requirement of computational effort for the whole process is not high thus outage location can be estimated in real time. The proposed method is straightforward and easy to implement. It also can be used for cross-checking line outage event via SCADA.

The rest of the paper is organized as follows. In Section II, the proposed line outage detection and localization method via

This work is supported by the CURENT Industry Partnership Program and the Engineering Research Center Program of the National Science Foundation and DOE under NSF Award Number EEC-1041877.

synchrophasor measurement are given. In Section III, the characteristic of power flow change redistribution during line outage is explored via PSS/E simulation on models of ISO-NE and TVA systems. The simulation study on the ISO-NE model is conducted in Section IV. The conclusions and future work are drawn in Section V.

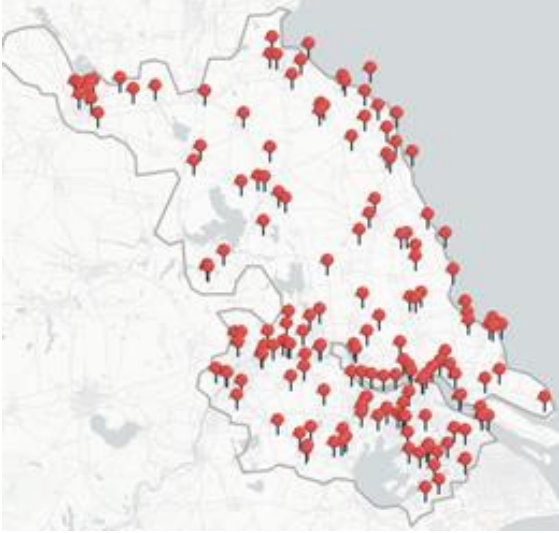

Fig. 1. Map of synchrophasor deployment in Jiangsu power grid [29]

## II. LINE OUTAGE DETECTION AND LOCALIZATION

The proposed method for line outage detection and location estimation in the paper includes two steps: (1). Line outage event detection using frequency measurement; (2). Outage line localization using active power measurement.

The frequency measurements from deployed PMUs is used to monitor that if there is ongoing a line outage event in the system. The principle of outage detection can be found in Ref. [34]-[36]. The frequency measurements are first fed into a moving median filter to remove random spikes and high-frequency noises. After that, a moving mean filer is used to extract the frequency trend as a reference. Then de-trended frequency data is subtracted frequency trend from filtered frequency measurements. Two thresholds are empirically set based on statistical analysis on historical data [34]. The event outage will be triggered using threshold evaluation. The event time can be determined via GPS timestamp on the measurements [37]-[40], which will be utilized further for event location estimation with active power measurements.

When a line outage event happens, the active power flow will be redistributed partially since the power flow on the tripped line has to transfer the rest of the system abruptly, which provides useful information for line outage location estimation. The Power Transfer Distribution Factors (PDTFs) of line $l$ respect to a power flow transaction $\Delta w$ in a lossless model is defined as [41]-[44]:

$$\varphi_l = \frac{\Delta l}{\Delta w} \quad (1)$$

where $\Delta w$ is MW of power transfer between two location and $\Delta l$ is the power transfer via branch $l$ respect to the transaction. Then, for an outage at line $m$, Line Outage Distribution Factor (LODF) is defined as the portion of pre-outage real power flow transfer to a line $k$ [41], which can be represented as

$$\varsigma_m = \frac{\Delta P_k}{P_m} = \frac{\varphi_k}{1-\varphi_m} \quad (2)$$

where $\Delta P_k$ is the power flow transfer changes at line $k$ and $P_m$ is the of pre-outage real power flow at line $m$. According to the definition in Eq.(1) and (2), PDTF and LODFs are less than 1.

Meanwhile, bus in pre-and post-outage condition must follow Gustav Kirchhoff's Current Law (KCL). Defining the power flow change at terminal $k_1$ is $\Delta P_{k_1}$, the power flow change outage follows:

$$\Delta P_{k_1} = \text{Sum}(\Delta P_j), \quad (3)$$
$$s.t., j \in branch\ connected\ to\ k_1$$

where $j$ is the indexes of the lines.

In an actual power system, the disturbance usually spread out from the source to the rest of the system. As a result, the bus with relatively large power flow changes might be closer to the location of the outage line, that is $\Delta P_{k_1} > \Delta P_{j_1}$ when distance of $k_1$ to the outage location is smaller than $j_1$. Therefore, the bus of the outage line is likely to have the largest power change in the system.

Using the active power change from synchrophasor, the location of outage can be estimated. Specifically, once a line outage event is detected, noise in active power measurement is filtered by a median filter. With detected event timestamp, the active power change between pre-outage and the post-outage is calculated with the filtered active power measurements. By ranking the active power change on the monitored transmission lines, the location of the outage line can be determined with maximum value. The process of the line outage detection and localization method is presented in Fig. 2.

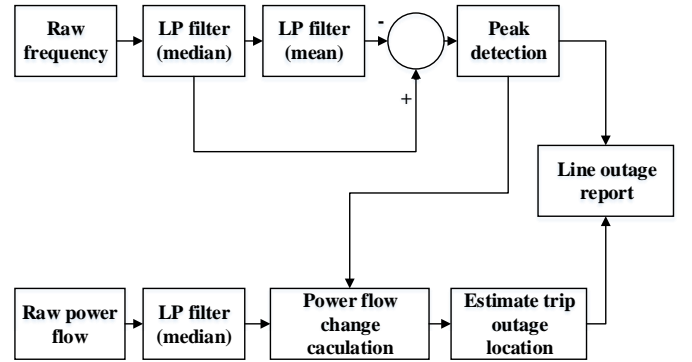

Fig. 2. Flowchart of line outage detection and localization

## III. DISTRIBUTION OF POWER CHANGE IN TVA AND ISO-NE

This section investigates the characteristic of power flow change distribution caused by line outage via PSS/E simulation. Line outage events are simulated in both ISO-NE and TVA systems, respectively. ISO-NE system consists of 3447 buses and 2479 branches in 71,992 mile². The total generation is 18.1GW and total load is 21.8 GW in the system. There are 16 tie lines, which carry 3.7 GW power flow, connecting to the system. TVA system has 1920 buses and 1720 branches. There are 28.1 GW generation and 31.6 GW load within TVA system. The TVA system are connected with external system via 70 tie lines and total 3.5 GW energy are delivered by the tie lies. The simplified system diagram of ISO-NE and TVA systems are showed in Fig. 3 and Fig. 4, respectively.

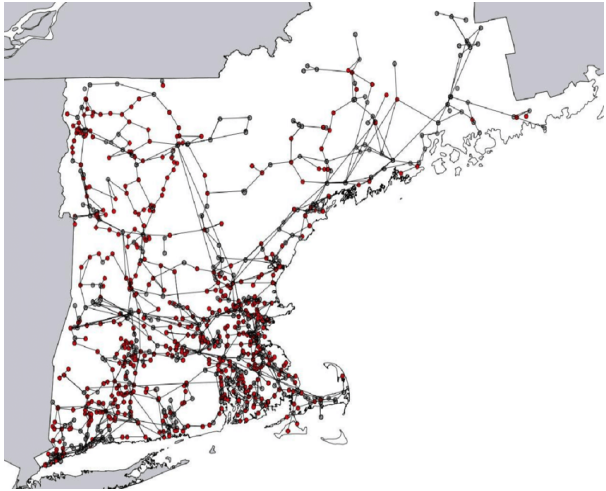

Fig. 3. ISO-NE model—transmission network map [45]

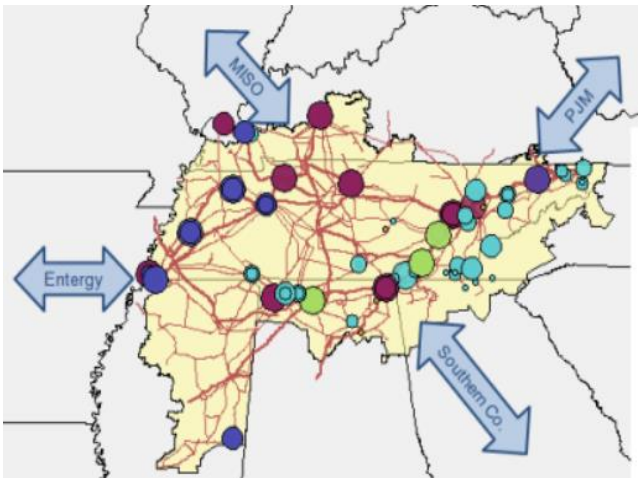

Fig. 4. Diagram for Tennessee Valley Authority [46]

TABLE I. ERROR ANALYSIS FOR LINE OUTAGE LOCATION ESTIMATION

| System | Case name | Termial$_1$ (Lat,Lon) | Termial$_2$ (Lat,Lon) | Power flow (MW) | Voltage level(kV) |
|---|---|---|---|---|---|
| ISO-NE | Line$_1$ | 41.51,-72.56 | 41.29,-72.90 | 407.36 | 345 |
| | Line$_2$ | 42.63,-71.05 | 42.70,-70.87 | 100.75 | 115 |
| TVA | Line$_3$ | 35.10,-85.02 | 34.05,-85.08 | 856.78 | 500 |
| | Line$_4$ | 37.78,-86.48 | 37.26,-86.98 | 246.72 | 161 |

For the purpose of a comprehensive study, 4 transmission lines in ISO-NE and TVA system are selected and tripped. The voltages levels of the outage line are from 115 kV to 500 kV. The terminal locations (latitude and longitude) of outage line and pre-outage real power flow on the lines are given in Table I. The locations of the disturbance and distribution of power flow changes caused by the line outage are shown from Fig. 5 to Fig. 6. It can be seen that the power flow changes at the terminals of the outage line have the highest value for the both cases. What is more, the power flow change closer the outage line generally has a larger value than the line far away from the event location.

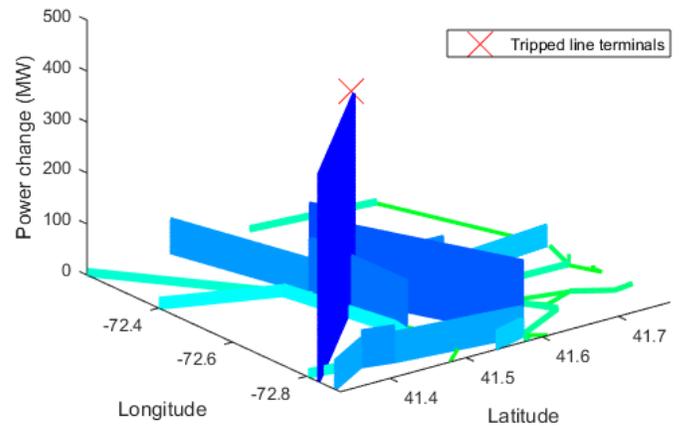

(a)

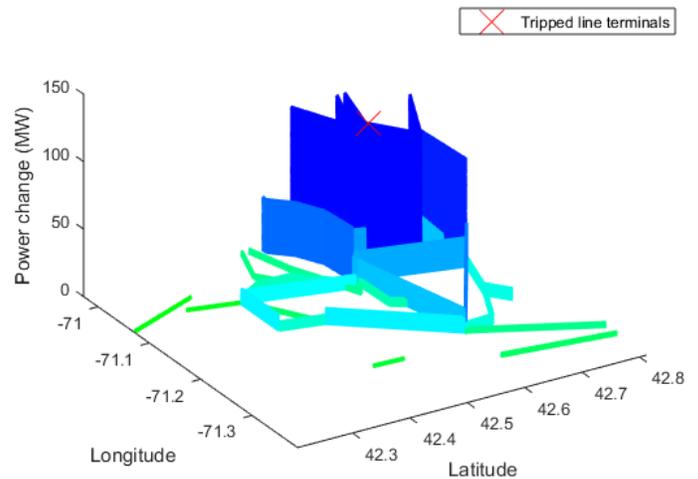

(b)

Fig. 5. Distribution of power change in ISO-NE system (a) Line$_1$ outage (b) Line$_2$ outage

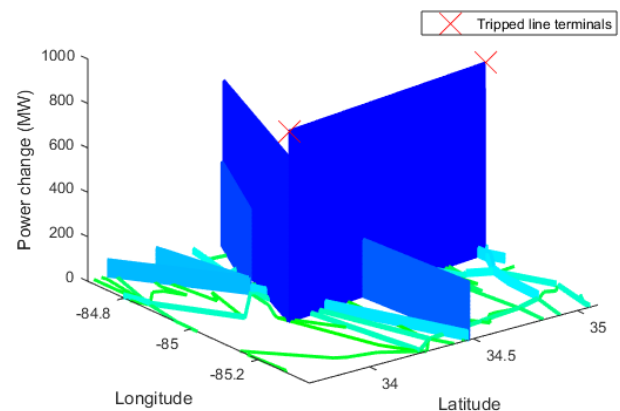

(a)

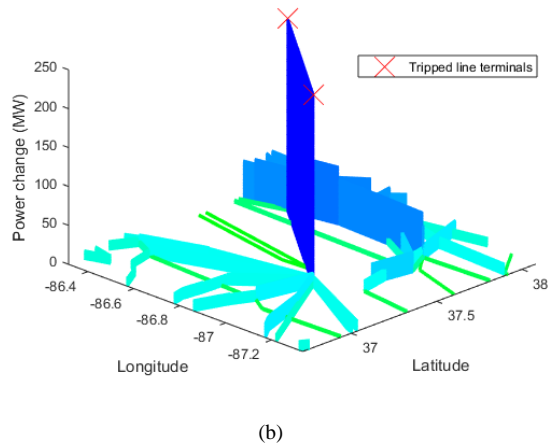

(b)

Fig. 6. Distribution of power change in TVA system (a) Line$_3$ outage (b) Line$_4$ outage

## IV. SIMULATION STUDY

To verify the effectiveness of the proposed method for line outage detection and localization, a simulation is conducted in ISO-NE which assumes that synchrophasor covers all 345 kV and part of (26%) 230 kV transmission lines. Line outages events are triggered in PSS/E to evaluate the performance of the events at different voltage levels. The parameters of the filter for line outage detection are selected based on [34], which are listed in Table II. The event detection module is implemented in C#, while location module is developed by MATLAB. The simulation tests are on a computer running a 64-bit Windows 10, with a 3.60 GHz Intel I7-7700U CPU and 16 GB memory.

TABLE II. PARAMETER SELECTION FOR LINE OUTAGE DETECTION

| Parameters | Values |
| --- | --- |
| Median filter window | 7 points |
| Mean filter size | 31 points |
| Detection window | 20 points |
| First peak threshold | 0.0045 Hz |
| Second peak threshold | 0.0025 Hz |

The line outage events can be successfully detected, and event time can be accurately recorded for all simulation cases. The location estimation of the events is further analyzed. The information of the line outage location and estimation error are in Table III. As shown in the Table III, the proposed method is able to identify the outage line location precisely, when the outage lines are monitored by synchrophasor. For the outage lines without synchrophasor monitoring (115 kV), the estimated location is close to actual outage line terminals. For the cases with a large error in 115 kV case, the actual outage lines are far away from PMU locations and the reported PMU is the closest location to the actual outage line terminals.

TABLE III. RESULT OF LINE OUTAGE LOCATION ESTIMATION

| Voltage level | Monitored by PMU | Cases numbers | Max error (Mile) | Average error (Mile) |
| --- | --- | --- | --- | --- |
| 345 kV | Y | 37 | 0 | 0 |
| 230 kV | Y | 20 | 0 | 0 |
| 230 kV | N | 8 | 13.72 | 6.32 |
| 115 kV | N | 30 | 82.49 | 10.42 |

TABLE IV. PERFORMANCE COMPARISON FOR LOCATION ESTIMATION

| Case name | Voltage level | Monitored by PMU | Power flow (MW) | Power change (mile) | Max freq. (mile) |
| --- | --- | --- | --- | --- | --- |
| 1 | 345 kV | Y | 725.17 | 0 | 93 |
| 2 | 230 kV | Y | 224.43 | 0 | 123 |
| 3 | 230 kV | N | 285.65 | 9.366 | 126 |
| 4 | 115 kV | N | 100.45 | 0 | 18 |

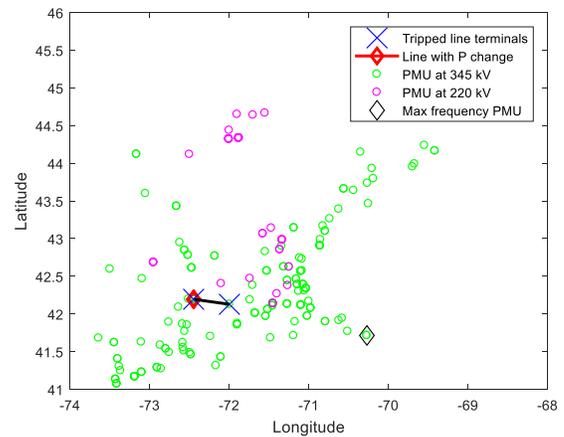

Fig. 7. Comparison of line outage localization (case 1)

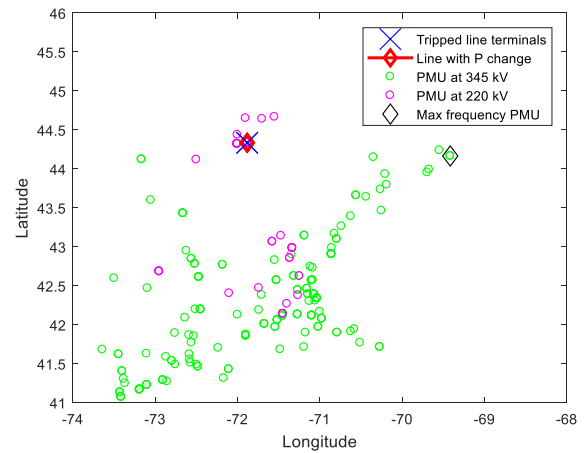

Fig. 8. Comparison of line outage localization (case 2)

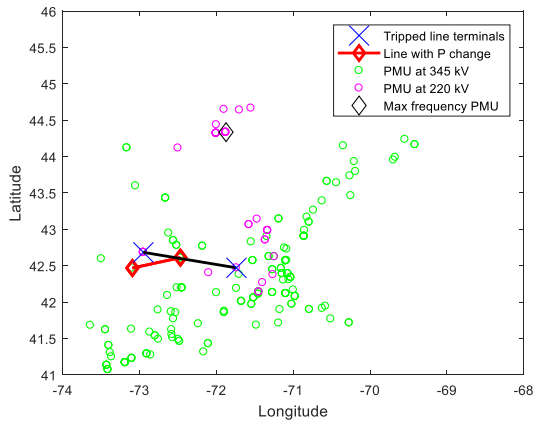

Fig. 9. Comparison of line outage localization (case 3)

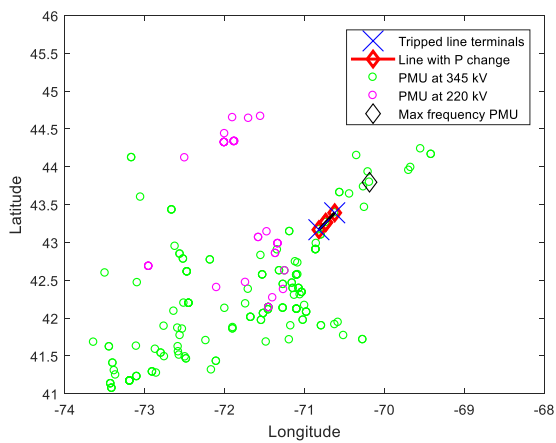

Fig. 10. Comparison of line outage localization (case 4)

For the purpose of comparison, the locations of four line outage cases from 115 kV and 345 kV are estimated by the proposed methods and the traditional methods using the maximum frequency magnitude change in Ref.[34] and [47]. The estimation errors for each case are given in Table IV. The estimation locations and actual line outage for each case are plotted from Fig. 7 to Fig. 10. As shown in these figures, distances between the estimated locations by proposed methods and the actual location of outage line are small while the estimated location by methods using frequency changes has significant deviations from actual outage locations.

## V. Conclusion and Future Works

Awareness of line outage event and its location is critical to prevent cascading outages in today's modern power system. This paper presents a fast line outage detection and localization method utilizing synchrophasor measurements. The line outage is first detected via a peak detector on synchronized frequency measurements, and location of the fault line is directly estimated via active power flow change. The proposed is straightforward and does not need the pre-knowledge on system topology and parameters. The feature of active power change distribution caused by line outage is explored in both TVA and ISO-NE system. A comprehensive simulation study in ISO-NE shows the method can precisely identify the outage line with reasonable accuracy. It can works as an effective tool for real-time line outage detection and localization.

Simulation results manifest that the proposed approach is promising for line outage detection and localization in large-scale power system. The performances of the proposed approach have not been validated with line outage events from real power grid. Additionally, the approach has not been fully tested for real-time implementation. Following are some future works for further development:

(1) Validate the proposed approached with confirmed line outage event from power grid utilities.

(2) Test the robustness of the approach with synchrophasor measurements with low quality data.

(3) Develop interface with available PMU data platform and evaluate the performance of this approach with real synchrophasor measurements